\documentclass[%
 aip,
 jmp,%
 amsmath,amssymb,
reprint,%
]{revtex4-1}

\begin{document}

\title{Quantum gravity, information theory and the CMB}


\author{Achim Kempf}


\affiliation{Departments of Applied Mathematics and Physics\\
           University of Waterloo, Waterloo, Ontario, Canada \\
           }


\maketitle

{\bf We review connections between the metric of spacetime and the quantum fluctuations of fields. In particular, we discuss the finding that the spacetime metric can be expressed entirely in terms of the 2-point correlators of the fluctuations of quantum fields. We also discuss the open question whether the knowledge of only the spectra of the quantum fluctuations of fields suffices to determine the spacetime metric. This question is of interest because spectra are geometric invariants and their quantization would, therefore, have the benefit of not requiring the modding out of the diffeomorphism group. Further, we discuss the fact that spacetime at the Planck scale need not necessarily be either discrete or continuous. Instead, results from information theory show that spacetime may be simultaneously discrete and continuous in the same way that information can. 
Finally, we review the finding that a covariant natural ultraviolet cutoff at the Planck scale implies a signature in the cosmic microwave background (CMB) that may become observable.}


\section{Introduction}

At any given point in history, the then known laws of Nature tell not only about the nature of Nature but also about the nature of us. The history of physics may, therefore, be viewed as a process of emancipation from our evolutionary heritage. 

For example, in early modern physics, the known laws of Nature were about subjects such as acoustics, optics and mechanics, directly corresponding to the set of senses that evolution happened to equip us with, such as vision, hearing and touch. 
In the meantime, the technology of experiments has superseded our sensory capabilities and, as a consequence, many `common sense' assumptions about Nature have been superseded by the discovery of much deeper laws of Nature. 

The unification of general relativity with quantum theory, see, e.g., \cite{kiefer}${}^{-}$\cite{sorkin} could be the last step towards fully uncovering the fundamental laws of Nature. As such it may then also be the last step of emancipation from our natural prejudices about Nature. 

From this perspective, what it will take to make way for the deepest understanding of Nature is not only the development of experimental technology that approaches the quantum gravity regime. It will also be necessary to identify the last, basic, human, common sense assumptions about Nature that need to be abandoned.   

For example, it appears to be common sense that spacetime provides a stage while particles and fields populate that stage. Perhaps, this duality of stage versus actors may ultimately need to be abandoned in favor of a unifying theory that makes no fundamental distinction between spacetime degrees of freedom and matter degrees of freedom, a theory in which that distinction only arises at low energies. 

In such a theory, conventional units, such as units of length or mass, units that were originally introduced for use on vegetable markets, could lose their operational meaning at Planck energies and may need to be replaced by more robust units, such as the bit and qubit, see, e.g.,  \cite{ak-qg-qc}${}^{-}$\cite{AK-Beethoven}. At present, due to insufficient experimental data, we can only speculate. 

What we do know, however, is that the unifying theory of quantum gravity will yield general relativity and quantum field theory as limiting cases. The mathematical framework of the unifying theory will, therefore, also naturally unify the mathematics of general relativity, i.e., differential geometry, with the mathematics of quantum theory, i.e., functional analysis.  

The present paper is to report on recent work with my group and collaborators in which we have explored connections  between  the  metric  of  spacetime  and  the quantum fluctuations of fields, including an approach to possibly identifying evidence of Planck scale physics in cosmological observations. The paper is based on an invited presentation at the Lemaitre Workshop, May 9-12, 2017 at the Vatican Observatory. 

\section{Spacetime could be simultaneously continuous and discrete in the same way that information can.}
As quantization literally means discretization, spacetime is often modeled as being ultimately a discrete structure \cite{rovelli}${}^{-}$\cite{sorkin}. Such models of spacetime are attractive, in particular, because they provide a natural ultraviolet cutoff. But they tend to come with the loss of local external symmetries such as translation and Lorentz invariance and it can also be difficult to obtain a smooth manifold of fixed dimension as a natural limiting case. By modeling spacetime instead as a continuous structure, such as a differentiable manifold, these problems can be avoided but then the challenge remains to appropriately model the expected phenomenon that quantum fluctuations of spacetime render the very notion of distance operationally ill defined at the Planck scale. 

Regarding the question of whether spacetime should be modeled as being continuous or discrete, it is interesting to note that these two possibilities are not mutually exclusive. There are examples of structures that are simultaneously continuous and discrete. In particular, in information theory, Shannon sampling theory establishes that information can be  simultaneously continuous and discrete. The same mathematics can be applied to physical fields, implying that  spacetime may be simultaneously continuous and discrete in mathematically the same way that information can \cite{AK-NJP}. 

To see this, let us begin by recalling the basic Shannon sampling theorem \cite{shannonsbook,thomascover}. The theorem concerns functions, $f(t)$, such as a signal, that are bandlimited, i.e., that contain only frequencies in a finite interval $(-\Omega,\Omega)$.  
Shannon's theorem holds that knowledge of the samples  $\{f(t_n)\}_{n=-\infty}^\infty$ of such a signal's amplitude that were taken at a spacing that obeys $t_{n+1}-t_n=(2\Omega)^{-1}$ is sufficient to reconstruct the signal at all times: 
\begin{equation}
    f(t) = \sum_n f(t_n)~\frac{\sin\left(2\Omega\pi(t-t_n)\right)}{2\Omega\pi(t-t_n)} ~~~~~~ \forall~t\in I\!\!R
\end{equation}
More generally, the signal can also be recovered from samples that are spaced irregularly, if the spacing of the samples is at most $(2\Omega)^{-1}$ on average in the Beurling sense \cite{zayed}${}^{-}$\cite{ferreira}. The price to pay when reconstructing a function from irregularly spaced samples is that the reconstruction becomes more sensitive to errors in the taking of the values of the samples. 

The theorem generalizes to functions over $I\!\!R^n$ and can then be applied to physical fields that possess an ultraviolet cutoff in the form of a bandlimitation which amounts to a hard momentum cutoff. For now, we let this be either a non-covariant 3-momentum cutoff or a covariant 4-momentum cutoff in a Wick-rotated theory with Euclidean signature. We will consider the case of manifolds with Lorentzian signature below.  
The value of the natural ultraviolet cutoff should presumably be chosen at the Planck scale of $10^{-35}m$, where quantum fluctuations of spacetime are thought to become strong. 

If spacetime possesses such a natural bandlimitation (which is non-covariant in three dimensions and covariant in the four-dimensional euclidean formulation) then fields and the equations of motion can be written on continuous spacetime and, equivalently on any spacetime lattice that possesses a minimum finite (e.g. Planckian) density of points. It is in this sense, via Shannon sampling theory, that spacetime could be described as being simultaneously continuous and discrete in the same way that the information in a bandlimited music signal is simultaneously continuous and discrete. 

The fact that the samples can be chosen irregularly spaced at the cost of reduced stability of the reconstruction of the function from the samples possesses implications when applied to quantum field theory where, in particular, it ensures the area law for entanglement entropy \cite{jason}. Bandlimitation is also closely related to generalized uncertainty principles 
\cite{GUP-a}${}^{-}$\cite{GUP-b} of the type
\begin{equation}
    \Delta x\Delta p \ge \frac{1}{2}\left(1+\beta (\Delta p)^2 + \dots \right)
\end{equation}
that arose from considerations of quantum gravity and string theory and that imply a finite lower bound on positions uncertainties, $\Delta x\ge \Delta x_{min}= \sqrt{\beta}$ (in units where $\hbar =1$). Indeed, these uncertainty principles
imply and  are implied by bandlimitation, with the minimum position uncertainty and the bandlimit being proportional \cite{withRTW}. This relationship between $\Delta x_{min}>0$ and bandlimitation also holds when $\Delta x_{min}$ is position dependent. 

The generalization to curved space is relatively straightforward \cite{ak-rtm}. To see this, let us recall that, for functions on the real line, bandlimitation is the restriction of the space of square integrable functions to the subspace spanned by the eigenfunctions $e^{ikx}$ of the self-adjoint derivative operator $D:=-i\partial_x$ whose eigenvalues $k$ obey $\vert k\vert <\Omega$. Equivalently, this is the restriction to the subspace spanned by the eigenfunctions of $-D^2$ whose eigenvalues, $\lambda$ obey $\lambda <\Omega^2$. On curved space, covariant bandlimitation is then the restriction of the Hilbert space of square integrable functions to the  subspace spanned by only those eigenfunctions of the Laplacian, $\Delta$, whose eigenvalues, $\lambda$ obey $\lambda <\Omega^2$. 

In this way, physical fields and their equations of motions can be written as living on any sufficiently dense spacetime lattice while, equivalently, they can also be written as living on continuous spacetime. As a consequence, if Nature possesses this ultraviolet cutoff, one obtains lattice regularizations while the external 
symmetries and conservation laws, such as those described by Killing vector fields, can be maintained. 

The presence of curvature on the manifold affects the sampling and reconstruction of fields in an interesting way. It has been shown \cite{ak-gilkey}, using a result of Gilkey \cite{gilkey,hawking}, that curvature locally modulates the bandwidth, i.e., that curvature modulates the density of degrees of freedom, in the sense of the optimal (most stable) density of sample points from which to reconstruct fields. It is an open question whether, vice versa, the local curvature and metric can be expressed entirely in terms of the modulation of the local density of the degrees of freedom of the matter fields. In this case,  curvature could be viewed as an information-theoretic concept based on a notion of local information carrying capacity \cite{AK-NJP,ak-gilkey}. 

An important feature that distinguishes\footnote{Technically, in Shannon's theorem, a bandlimited signal is assumed to obey Dirichlet boundary conditions in the Fourier domain. A conventional lattice theory implements periodic boundary conditions in the Fourier domain.}
a bandlimitation from a conventional lattice cutoff is that in the case of the bandlimitation the field does not just live on one lattice. Instead, democratically, there are infinitely many sampling lattices on which the same field can be represented and on each lattice of sufficiently dense spacing there is enough information in the field's amplitude samples to reconstruct the field everywhere. A bandlimited music signal that has been reconstructed from samples via Shannon's theorem does not retain any memory of the choice of sample lattice that it was reconstructed from.

It has been pointed out \cite{ak-lattices} that this freedom to choose any sampling lattice, among all sampling lattices of sufficiently tight average spacing, is mathematically linked to unitary groups that arise from von Neumann's theory of self-adjoint extensions. This invites speculation that the degrees of freedom that are cut off at the Planck scale re-emerge to form the internal degrees of freedom of the gauge principles of the standard model, with their unitary gauge groups. In this case, local isospinors might be, in a sense, `locating' field degrees of freedom within Planckian volumes, while the gauge principle expresses the non-observability of that localization.     

So far, many of these results and speculations are confined to curved spaces with Euclidean signature. This is because the generalization of Shannon sampling theory to Lorentzian manifolds is generally mathematically nontrivial as bandlimitation then amounts to cutting off the spectrum of the hyperbolic d'Alembertian instead of cutting off the spectrum of the Laplacian which is an elliptic operator. However, a number of results for the generalization of Shannon sampling theory to Lorentzian manifolds have been obtained, see, e.g., \cite{ak-rtm,aidan1,aidan2}. 

The physics of such a covariant cutoff on the spectrum of the d'Alembertian is best understood in the path integral formulation of quantum field theory. There, the path integral involves a formal sum over a space of field configurations. The special field configurations that are on the mass shell, i.e., that obey the classical equations of motion, contribute the classical solutions while the other field configurations describe quantum fluctuations away from the classical solutions. The covariant bandlimitation is implemented by restricting the space of field configurations that one sums over in the quantum field theoretic path integral to the subspace spanned by the eigenfunctions of the d'Alembertian whose eigenvalues are bounded (above and below) by the Planck scale. The cutoff, therefore, eliminates the most extreme quantum fluctuations. Technically, this means in perturbation theory that in loops of Feynman graphs, the masses of virtual particles are restricted to masses below the Planck mass. 

Let us now discuss what this type of Lorentz covariant natural ultraviolet cutoff implies for the short-distance structure of the theory. This question is pressing because it should be hard to reconcile the presence of a finite natural minimum length scale with relativity, given that any length can be Lorentz contracted by a boost. In fact \cite{ak-rtm} the natural ultraviolet cutoff though covariant bandlimitation generalizes Shannon sampling theory so that it provides a beautiful new way to reconcile the presence of a minimum length with relativity. 

The way that covariantized Shannon sampling theory reconciles the existence of a kind of minimum length with covariance is easiest to see in the case of Minkowski space. Here, after Fourier transforming, the covariant cutoff amounts to implementing the inequality
\begin{equation}
    |p_0^2-\vec{p}^2| < \Omega^2
\end{equation}
This equation does not imply an upper bound on the magnitude of the three-momentum $\vec{p}$. Therefore, arbitrarily short spatial wavelengths can still occur. However, the inequality implies that spatial modes with very large $\vec{p}$, i.e., spatial wavelengths that are significantly smaller than the Planck length, possess an exceedingly small bandwidth in time (i.e., the range of allowed $p_0$ values becomes exceedingly small), and these transplanckian wavelengths are, therefore, effectively frozen out as their dynamics becomes trivial. 

To summarize, the statement is that wavelengths shorter than the Planck length dynamically freeze out in the sense that their bandwidth in time becomes exceedingly small, which means that they are fully determined by samples in time which have an exceedingly large spacing, which implies that the dynamics of these wavelengths becomes simple and predictable, hence effectively frozen. The covariance of this statement is maintained under Lorentz transformations because the inverse of the temporal bandwidth time dilates correspondingly as the size of spatial wavelengths Lorentz contracts. Also, while every spatial mode obeys a corresponding conventional sampling theorem in time, vice versa, every temporal mode obeys a corresponding conventional sampling theorem in space. 

The generalization of Shannon sampling to cosmological spacetimes has  been developed in \cite{aidan1,aidan2}. In this case, each comoving spatial mode possesses a temporal sampling theorem. The shorter the proper wavelength of this mode, the smaller is this mode's effective bandwidth in time. This means that the number of sample points that need to be taken to capture a comoving mode's behavior from past infinity, or as the case may be from the big bang, to any point in the future can then be finite. In particular, if we follow modes back in time to when their proper wavelength drops below the Planck length we find that only on the order of one sample point is needed to capture the mode's behavior there. Vice versa, this means that as, during expansion, comoving modes keep crossing the Hubble length, they become unfrozen in the sense that, literally, their bandwidth becomes finite. In the next section we will discuss the application of generalized Shannon sampling to cosmic  inflation \cite{aidan1,aidan2}. The generalization of Shannon sampling to black holes is in progress.

\section{A covariant natural ultraviolet cutoff at the Planck scale implies a potentially observable signature in the CMB.}

Since the Hubble length during inflation was likely only about five orders of magnitude from the Planck length, it has been proposed to search for signatures of Planck scale physics in the cosmic microwave background (CMB), \cite{cosmo-a}${}^{-}$\cite{cosmo-b}. Such studies require a concrete model for how quantum gravity influences the quantum field theoretic framework a few orders of magnitude from the Planck scale. 
In the literature, this influence tends to be modeled through the introduction of either discretization, a minimum length, a generalized minimum length uncertainty principle or modified dispersion relations. In such models, it tends to be difficult to maintain covariance. Without covariance, however, it is difficult to distinguish how much of a predicted effect on inflation is due to quantum gravity, and how much is due merely to the breaking of covariance. In particular, no consensus has been reached regarding the crucial question whether the impact of quantum gravity on the CMB spectrum should be expected to be of first or second order in the key dimensionless ratio: 
\begin{equation}
\sigma ~=~ \frac{\mbox{Planck length} }{ \mbox{Hubble length during inflation}}  \end{equation}
If the effects are merely linear rather than of higher order in $\sigma$, then eventual experimental testability may not be ruled out.  

Recently, the covariant information-theoretic natural ultraviolet cutoff at the Planck scale that was discussed in the previous section has been implemented into the basic single-field inflationary scenario with DeSitter and power law expansion \cite{aidan1,aidan2}. This calculation maintains covariance 
and involves the application of von Neumann's theory of self-adjoint extensions to the d'Alembertian on expanding spacetimes. A benefit of employing von Neumann's formalism is that it yields a new powerful unified method for implementing initial conditions and the choice of vacuum (Bunch Davies) in the field propagator. 

The main result of these calculations is the prediction that the perturbation spectrum, as generated in the inflationary scenario, should be modulated with $k$-dependent small oscillations whose frequency is determined by the slope of the slow roll and whose amplitude is, crucially, linear in $\sigma$. It is not excluded, therefore that these characteristic oscillations may, therefore, become testable in the CMB. In principle, the only free parameter in these predictions is the value of the cutoff, for which the natural choice is the Planck scale. An open question is how the predictions would be affected if the natural ultraviolet cutoff on the spectrum of the d'Alembertian is not a sharp cutoff but a smooth tapering off. A plausible conjecture is that the predicted small oscillations on top of the main CMB spectrum would start to wash out.

\section{The metric can be deduced from knowledge of the correlator of vacuum fluctuations.}

The equations of motion of fields can be viewed as the equations of motion for coupled harmonic oscillators (with the usual caveats for fermionic fields and except for the non-harmonic oscillators of self-interacting fields). The fields of the various particle species are acting as driving forces for each other's oscillators. 

The spatial derivatives in the field equations couple local field oscillators to their neighbors so that local excitations spread in space. 
The coupling between neighboring local quantum field oscillators implies that the quantum fluctuations of neighboring field oscillators are correlated and entangled, and the more so the closer the oscillators are. As is well known, the relatively fast decay of this entanglement with growing distance, for dimensions larger than (1,1), means that the entanglement between the field degrees of freedom of the inside and the outside of a region is governed by the size of the dividing boundary and, therefore, tends to obey an area law. 

As was shown in \cite{siavash}, the spatial decay of the correlations between neighboring field oscillators in the vacuum state also reveals a direct connection between quantum vacuum fluctuations and spacetime curvature. To see this, recall that the wave operators, such as the d'Alembertian in the case of the Klein Gordon fields, 
contain metric-dependent coefficient functions and are, therefore, generally impacted by the presence of curvature. Spacetime curvature impacts the quantum ringing of fields in spacetime in a mathematically similar way to how the curvature of a glass vase affects the spectrum with which the vase can ring. Given that curvature impacts quantum field fluctuations, the question then arises to what extent knowledge of the quantum fluctuations' statistics can be used to reconstruct the metric. In fact, it has been shown that knowledge of a 2-point correlator of quantum fields, such as the Feynman propagator, $G(x,y)$, can be used to reconstruct the metric. In the case of the free Klein Gordon field in $n$-dimensional Lorentzian spacetimes, for $n>2$: 
\begin{equation}
    g_{\mu\nu}(x) = -\frac{1}{2}\left[\frac{\Gamma(n/2-1)}{4\pi^{n/2}}\right]^{\frac{2}{n-2}}\lim_{y\rightarrow x}\frac{\partial}{\partial x^\mu}\frac{\partial}{\partial x^\nu}~G(x,y)^{\frac{2}{2-n}}
\end{equation}
The original motivation for trying to derive the metric from the 2-point correlator has been the fact that knowledge of the 2-point correlator implies knowledge of the light cones because of the divergence of the correlator on the light cone, while, as is well known, \cite{HawkingEllis}, a Lorentzian manifold is determined up to a conformal prefactor by knowledge of its light cones. The question that remained was whether or not the Feynman propagator also encodes the conformal factor.

There is a simple reason for why it does and why, therefore, it is  possible to derive the tensorial metric from a bi-scalar 2-point correlator. The reason is that the 2-point correlator decays with the invariant distance of the events $x$ and $y$, and it can, therefore, serve as a proxy for distance measurements. And to know infinitesimal invariant distances, in this case by means of the propagator, is to know the metric.  
In further work, \cite{yasaman}, it has been shown that also within the framework of causal set theory, the propagator carries the complete information about the discretized spacetime, i.e., any causal set can be re-constructed from knowledge of a propagator. This confirms that a propagator on causal sets does not only contain the information about the light cone structure of the spacetime but that it does also contain the information about the spacetime's conformal factor, which is information that in the case of causal sets is normally encoded in the density of the sprinkled events. 

\section{Towards describing spacetime in terms of geometric invariants}

In the previous section, we discussed that knowledge of the propagator, $G(x,y)$ of, for example, a free Klein Gordon field allows one to reconstruct the metric $g_{\mu\nu}(x)$ and, therefore, to obtain the Lorentzian manifold of the underlying the spacetime. In this way, the metric and curvature can be expressed entirely in terms of the correlations of quantum vacuum fluctuations of a field that lives on the spacetime. 

This description of the metric of spacetime still suffers from the fact that it is a highly redundant description, because of its variance with the diffeomorphism group of changes of coordinates. This matters if we are to try to quantize gravity by path integrating over a set of spacetimes. In the path integral, each spacetime should occur only once, similar to how, in a gauge theory, only one field should occur out of each gauge equivalence class. Technically, a spacetime is a Lorentzian structure, i.e., an equivalence class of differentiable manifolds with Lorentzian metric tensors that are connected by isometric diffeomorphisms. Since it is difficult to gauge fix in the case of gravity, it would be desirable to find a description of spacetimes, i.e., of Lorentzian structures in terms of only geometric invariants, i.e., a description that is diffeomorphism invariant.  

From this perspective, let us reconsider the fact that knowledge of the propagator constitutes a complete description of the spacetime. The propagator, as a function, is dependent on the choice of coordinate system, and, therefore, the propagator provides a coordinate system dependent description of a spacetime. This suggests to ask if there is information in the propagator which is coordinate system independent, how this  information might be obtained, and whether it is sufficient to obtain a complete description of the spacetime, in terms of geometric invariants. 

Given that the spectrum of the 2-point correlator when viewed as an operator on functions on the manifold is independent of the choice of basis in that function space, it is natural to ask if the spectrum of the correlator determines the metric of the underlying spacetime. We know that the correlator does entirely determine the spacetime metric, but also that it is a redundant description because it is variant with the choice of coordinate system. The spectrum of the quantum vacuum fluctuations, i.e., the quantum noise on the spacetime, on the other hand, consists of geometric invariants but it may represent merely a subset of the geometric information contained in the correlator. A priori, the information contained in just the spectrum of the correlator may or may not be too small to determine the metric fully.   

To address this question, let us now retreat to the simpler case of euclidean spacetimes of finite volumes, technically, compact Riemannian manifolds. In this case, the spectrum of the 2-point correlator is discrete and it is the inverse of the Laplace operator's spectrum (while in Lorentzian spacetimes the propagator is a right inverse of the d'Alembertian, the propagator is not self-adjoint due to its anti-Hermitean component that is in the kernel of the d'Alembertian). The spectra of the 2-point correlator and the Laplacian are coordinate system invariant, i.e., they are geometric invariants. The question is how much information about the metric is contained in the spectrum of the Laplacian. 

We have arrived, from a new perspective, at a central question of the field of spectral geometry: to what extent does the spectrum of the Laplacian (on a compact Riemannian manifold) determine the underlying manifold.   

This is a hard problem. First, let us discuss why it is that the eigenvalues of the correlator (or Laplacian) may contain strictly less information than the correlator itself. To this end, let us consider that changes of coordinate system imply a change of basis in the Hilbert space of functions on which the 2-point correlator acts as an operator. Since the eigenvalues of the correlator are basis independent, the spectrum is coordinate system independent. Crucially, this does not imply the converse. While it is true that the eigenvalues of the correlator are independent of the basis in the Hilbert space, not every change of basis in the Hilbert space of square integrable functions over the manifold arises from a change of coordinate system. This means that by retaining, among all the information in the correlator, only its eigenvalues we are retaining only the information in the correlator that is invariant under \it all \rm changes of basis in the Hilbert space of functions on the manifold. This may be too little geometric information. There could be information about the metric in the correlator which is invariant under all those changes of basis in the Hilbert space that are induced by changes of coordinates but that is not invariant under \it all \rm changes of basis in the Hilbert space.   

This distinction matters, at least in dimensions larger than 2 because in dimensions larger than two, the spectrum of the Laplacian cannot determine the metric fully. Counter examples that prove this point have long been known, such as certain high-dimensional tori, and such as those examples that are generated by Sunada's method. For a review, see, e.g., \cite{datchevhezari}. 

But there is also a relatively new systematic way to probe the amount of metric information contained in the spectrum of the Laplacian as a function of the dimension \cite{AK-NJP,ak-aasen}. This method shows that the spectrum of the Laplacian in a precise sense contains sufficient information in the case of two dimensions while confirming that the spectrum does not suffice in higher dimensions. The new method also shows that if the spectra of certain other Laplace operators are considered, they may provide sufficient information to determine the metric, even in higher dimensions. 

The key idea of this method is to simplify the highly nonlinear problem of spectral geometry, i.e., the problem of determining a metric from a spectrum, by linearizing the problem. To this end, the problem of determining a metric from a spectrum is replaced by the problem of determining an infinitesimal change of a known metric from the corresponding infinitesimal change of its known spectrum. Technically, this amounts to taking the nonlinear map from metrics to spectra, calculating  the tangent map and checking for its invertibility. 
If successful, such infinitesimal reconstruction steps can then be iterated and effectively integrated to obtain finite changes of shape from finite changes of the spectrum. 

This program has been carried out for 2-dimensional compact Riemannian manifolds \cite{ak-aasen,panine1,panine2}. In this case, the metric, being a 4-component symmetric matrix, has three independent matrix elements of which two can be fixed by choosing two new coordinates, thus leaving effectively only one degree of freedom. This scalar function can be expanded in the eigenbasis of the Laplacian and is therefore describable by its discrete set of coefficients $\phi_n$ with respect to the Laplacian's eigenbasis. The spectrum of the Laplacian is given by the discrete set of eigenvalues $\lambda_n$. Let us now impose an ultraviolet cutoff by cutting off the Laplacian at some, for example, Planckian value $\Omega^2$, i.e., we restrict the function space to that finite-dimensional function space of dimension say $N$, which is spanned by the eigenfunctions of the Laplacian to eigenvalues obeying $\lambda <\Omega^2$. Intuitively, this is the situation of finite vision and hearing, where we cannot see ripples in the manifold shorter than the scale set by $\Omega^2$, nor can we hear the associated eigenfrequencies above the scale $\Omega^2$. 

Important here is the fact that there are in this case as many eigenvectors as there are eigenvalues (allowing for degeneracies). This means that there are as many coefficients describing the metric as there are coefficients describing the spectrum. The linear matrix that maps infinitesimal changes in the metric, i.e., infinitesimal changes in the $N$ coefficients $\phi_n$ into the infinitesimal changes in the eigenvalues, $\lambda_n$, is, therefore, a quadratic matrix. Generically, its determinant is nonzero and it is, therefore, invertible. The calculation of changes of the metric from changes of the  spectrum should, therefore, generally work, at least infinitesimally, in two dimensions. This has also been demonstrated explicitly numerically \cite{ak-aasen}. 

This consideration also shows why there is an obstruction in dimensions higher than 2. The reason is that in higher dimensions the metric is no longer effectively scalar because it possesses more than one degree of freedom. But only scalar perturbations of the metric can be expanded in the eigenbasis of the Laplacian on scalars. Therefore, changes in the spectrum of the  Laplacian on scalars then cannot contain all information about changes in the metric. But this observation also suggests a way forward. The perturbations of the metric in any dimension can be expanded in the eigenbasis of any Laplacian operator that acts on covariant symmetric 2-tensors because its eigenvectors will span that function space. In this way, it should be possible to obtain again a square tangent matrix after implementing an ultraviolet cutoff. 

As discussed in the introduction, the sought-after theory of quantum gravity will need to contain general relativity and quantum field theory on curved spacetime as limiting cases. The theory of quantum gravity will, therefore, have to naturally combine their mathematical frameworks, differential geometry and functional analysis. In order to  explore the mathematical ways in which this can happen, it will be interesting to further develop spectral geometry, Shannon sampling theory and their information-theoretic implications for Lorentzian manifolds. 

\bf Acknowledgement: \rm
This work has been supported by the Discovery Program of the National Science and Engineering Research Council of Canada (NSERC). 




\end{document}